\documentclass[aps,prd,onecolumn,groupedaddress,preprintnumbers,superscriptaddress,showpacs,nofootinbib]{revtex4}
\usepackage{amsmath,amssymb,latexsym}
\usepackage{subfigure}
\usepackage[dvips]{graphicx}
\usepackage{color}

\begin{document}

\title{
Hawking temperature for near-equilibrium black holes
}


\author{Shunichiro Kinoshita}
\email{kinoshita_at_tap.scphys.kyoto-u.ac.jp}
\affiliation{   
Department of Physics, Kyoto University, 
Kitashirakawa Oiwake-Cho, 606-8502 Kyoto, Japan
}
\author{Norihiro Tanahashi}
\email{tanahashi_at_ms.physics.ucdavis.edu}
\affiliation{
Department of Physics, University of California, Davis, California 95616, USA
}

\preprint{KUNS-2372}

\newcommand{\uzero}{u_*^{(0)}}
\newcommand{\uone}{u_*^{(1)}}

\date{\today}

\begin{abstract}
 We discuss the Hawking temperature of near-equilibrium black holes using a 
 semiclassical analysis.
 We introduce a useful expansion method for slowly evolving spacetime, and
 evaluate the Bogoliubov coefficients using the saddle point approximation. 
 For a spacetime whose evolution is sufficiently slow,
 such as a black hole with slowly changing mass, 
 we find that the temperature is determined by the surface gravity
 of the past horizon.
 As an example of applications of these results, we study the Hawking temperature
 of black holes with
 null shell accretion in asymptotically flat space and the AdS--Vaidya spacetime.
 We discuss implications of our results in the context of the AdS/CFT correspondence.
\end{abstract}

\maketitle

 \section{Introduction}
 
 The fact that black holes possess thermodynamic properties has been
 intriguing in 
 gravitational and quantum theories, and is still attracting interest.
 Nowadays it is well-known that black holes will emit thermal radiation with
 Hawking temperature proportional to the surface gravity of the event horizon~\cite{Hawking:1974sw}.
 This result plays a significant role in black hole thermodynamics.
 Moreover, the AdS/CFT correspondence~\cite{Maldacena:1997re} opened up new insights about
 thermodynamic properties of black holes.  
 In this context we expect that thermodynamic
 properties of conformal field theory (CFT) matter on the boundary would respect those of black
 holes in the bulk.
 In particular, stationary black holes correspond to thermal equilibrium 
 of the dual field theory at finite temperature equal to the bulk Hawking
 temperature~\cite{Witten:1998zw}.

 It is interesting to study if such thermal properties persist
 when time dependence is turned on.
 From a technical viewpoint, the original derivation of the Hawking 
 radiation~\cite{Hawking:1974sw} was performed on a static or stationary
 background (precisely speaking, as an approximation at late time), 
 and it is not obvious how to accommodate dynamical spacetimes into the scheme. 
 From a physical viewpoint, such a study can be useful since realistic black 
 holes in our universe are dynamical or at least quasistationary. For example, 
 processes like black hole formation after a gravitational collapse or
 black hole evaporation due to the Hawking radiation may involve highly
 time-dependent phases, and it is not obvious how the Hawking radiation will 
 behave in such systems.

 When time dependence is sufficiently weak, however, we may expect thermodynamic 
 properties of a black hole to persist.
 This expectation partially stems from physical insight on ordinary
 thermodynamics systems, in which thermodynamic properties persists when the
 system is sufficiently near equilibrium and quasistationary.
 If black hole thermodynamics is robust enough, such insight on usual
 thermodynamic systems leads to the above expectation on dynamical spacetimes and
 black holes.
 Many studies have been made on generalizing Hawking temperature for
 dynamical black hole
 approaches~\cite{Harada:2000ar,Saida:2007ru,Nielsen:2007ac,Hayward:2008jq,Barcelo:2010pj,Barcelo:2010xk,Mazumder:2011gk},
 and they support such an expectation.

 Within the context of the AdS/CFT correspondence, thermodynamic properties of
 dynamical black holes lead to new insights into the dynamics of
 strongly coupled 
 field theory on the boundary. 
 One typical problem in this line is thermalization processes in the boundary
 field theory, which is holographically modeled by formation of the  bulk black hole
 and its equilibration into a stationary state.
 Much effort is devoted to such problems in connection with,
 for example, application of the AdS/CFT correspondence to QCD physics like that in 
 the RHIC experiment, or to nonequilibrium phenomena in condensed matter physics and
 fluid dynamics~\cite{Son:2002sd,Janik:2006gp,Kinoshita:2008dq,Chesler:2008hg,Murata:2010dx,Das:2010yw,Hubeny:2010ry,Ebrahim:2010ra,Hashimoto:2010wv,Erdmenger:2011jb,CaronHuot:2011dr,Balasubramanian:2011ur,Garfinkle:2011hm}.
 In this context, the Hawking radiation in the bulk is interpreted as quantum 
 fluctuation in the boundary theory, and it plays an important role in certain 
 setups~\cite{deBoer:2008gu,Aharony:2005bm}. 

 Based on these interests,
 in this paper we consider nonstationary spacetimes, and
 try to determine the Hawking temperature measured by observers 
 in the asymptotic region.
 One naive way to do this would be to define a certain time coordinate to associate 
 temperature of the black hole (future) horizon to that measured by asymptotic observers.
 Even though this procedure is seemingly natural from the viewpoint of black hole 
 thermodynamics or horizon dynamics, 
 it is puzzling from the viewpoint of causality, since this temperature
 is determined using information in the future region for that observer.
 In the AdS/CFT setup,
 if we were to define temperature at a point on the 
 boundary in this way, we would need information in the bulk region, which is not 
 causally accessible from that boundary point. 
 Such a property would not be desirable
 if we care about causality in the bulk and boundary,
 especially when time dependence comes into the story.

 To avoid such problems, we 
 consider the conventional derivation of the Hawking
 radiation based on Bogoliubov transformations for nonstationary
 background.
 An advantage of this method is that we can naturally determine 
 temperature measured by asymptotic observers using only information 
 causally available to them.%
 \footnote{
 The essential part of the calculation is similar to Ref.~\cite{Barcelo:2010xk}.
 }
 Particularly, we focus our attention on an eternal black hole
 rather than black hole formation by gravitational collapse because we
 are interested in near-equilibrium system such as transition from an
 equilibrium state to another equilibrium state.%
 \footnote{
 The derivations are quite similar even on the background spacetime with gravitational
 collapse and black hole formation. See also comments in Sec.~\ref{Sec:summary}.
 }
 In other words, we focus on late-time dynamics after the black hole 
 has formed.
 We clarify how the Hawking temperature
 changes when a black hole becomes dynamical, for instance, due to mass
 accretion to the black hole,
 and show that the Hawking temperature of dynamical black holes can be 
 naturally associated with ``surface gravity'' of the past horizon.
 
 The paper is organized as follows.
 In Sec.~\ref{sec:Bogoliubov_coeff}, we evaluate the Bogoliubov coefficients using the
 saddle point approximation and show that the ``surface gravity'' of the
 past horizon gives the Hawking temperature observed in the asymptotic region.
 In Sec.~\ref{sec:applications},
 as illustrations of the use of our method,
 we consider applications to simple examples in asymptotically flat
 and AdS cases. 
 We summarize and discuss implications of the results in Sec.~\ref{Sec:summary}.

 \section{Estimation of Bogoliubov coefficients}
 \label{sec:Bogoliubov_coeff}

 We estimate Bogoliubov coefficients in order to
 define Hawking temperature for a nonstationary background in this section.
 We consider an eternal black hole in thermal equilibrium 
 with its surroundings, that is, in the Hartle--Hawking state.
 In asymptotically flat cases, we should immerse the black hole into a
 thermal bath to realize this state.
 In asymptotically AdS cases, on the other hand, it is naturally realized
 due to its boundary condition~\cite{Klemm:1998bb,Hemming:2000as}
 because the black hole is enclosed in the AdS boundary. 

 We exploit the saddle point approximation to evaluate the Bogoliubov
 coefficients and determine the temperature from them.
 After reviewing usage of the approximation for the static spacetime in
 Sec.~\ref{static}, we consider its extension to nonstationary spacetime in
 Sec.~\ref{dynamical}. In this extension to nonstationary spacetime,
 we introduce a quantity, $\kappa(u)$,
 which may be interpreted as an extension of the surface gravity of static
 Killing horizon.
 In Sec.~\ref{past},
 we reinterpret this quantity from a geometric point of view,
 and clarify that $\kappa(u)$ is naturally associated with the past horizon of
 the eternal black hole.
 
  \subsection{Static spacetime and the saddle point approximation}
  \label{static}
  
  For simplicity, we consider the two-dimensional part of the spacetime
  consisting of time
  and radial directions, that is, we focus on the s-wave sector of radiation.
  We also adopt the geometric optics approximation, and
  neglect the backscattering of the waves due to the curvature of spacetime.
  We introduce the null coordinate $u$ which gives a natural time for
  observers in the asymptotic region (null infinity). 
  If the spacetime is stationary, this time corresponds to the Killing time.
  We introduce another null coordinate $U$ which is the affine parameter
  on the past horizon.
  For the stationary case it becomes the familiar Kruskal coordinate.
  Because lines described by $u=\text{const.}$ are outgoing null geodesics, 
  geodesic equations give a relation $U=U(u)$ between the two null coordinates.
  
  Now, we consider a massless scalar field.\footnote{
  In asymptotically AdS cases we should consider a conformally coupled
  scalar field.}
  In the current setup, solutions of the field equation are simply given
  by arbitrary functions of each null coordinate.
  In a standard manner we can define positive frequency modes with
  respect to $u$ and $U$, respectively.
  Then, the Bogoliubov transformation between two sets of modes
  $u_\omega \propto e^{-i\omega u}$ and
  $\bar u_{\hat{\omega}} \propto e^{-i\hat{\omega} U}$ is determined by the Bogoliubov coefficients 
  \begin{equation}
   \left.
    \begin{aligned}
     \alpha_{\omega\hat{\omega}}\\
     \beta_{\omega\hat{\omega}}
    \end{aligned}
	     \right\} = \pm \frac{i}{2\pi}\sqrt{\frac{\hat{\omega}}{\omega}}
   \int^{\infty}_{-\infty} \mathrm du \frac{\mathrm dU}{\mathrm du}
   e^{\pm i \omega u - i \hat{\omega} U(u)},
   \label{eq:alpha_beta}
  \end{equation}
  where the upper and lower signs correspond to
  $\alpha_{\omega\hat{\omega}}$ and $\beta_{\omega\hat{\omega}}$,
  respectively.

  Now, to evaluate those coefficients by the saddle point approximation,
  we will consider the integral 
  \begin{equation}
   \int^{\infty}_{-\infty} \exp\phi(u)\mathrm du,\label{eq:integral}
  \end{equation}
  where 
  \begin{equation}
   \phi(u) \equiv 
    \log\frac{\mathrm dU}{\mathrm du} \pm i \omega u - i \hat{\omega} U(u).
    \label{phi}
  \end{equation}
  Saddle points are located at $u=u_*$ given by 
  $\phi'(u_*) = 0$, where 
  \begin{equation}
  \phi'(u)
   = \frac{\mathrm d}{\mathrm du}\log\frac{\mathrm dU}{\mathrm du}
   \pm i \omega - i \hat{\omega} \frac{\mathrm dU}{\mathrm du},
   \label{phip}
  \end{equation}
  and a prime denotes a derivative.

  Consider a static black hole, for example.
  When the surface gravity at the Killing horizon is $\kappa$, the relation
  between the two coordinates $u$ and $U$ is given by 
  $U(u) = - \exp(-\kappa u)$, which is the coordinate transformation for 
  the Kruskal coordinate.
  Then, we have 
  \begin{equation}
   \phi'(u_*)
    = - \kappa 
    \pm i \omega - i \hat{\omega} \kappa e^{-\kappa u_*} = 0,
    \label{eqsaddle}
  \end{equation}
  and the saddle point 
  \begin{equation}
   u_* = 
    -\kappa^{-1}\log\left(\frac{ire^{\mp i\theta}}{\hat{\omega}}\right)
    = - \frac{1}{\kappa}\log\frac{r}{\hat{\omega}}
    - \frac{i}{\kappa}\left(\frac{\pi}{2} \mp \theta\right),
  \end{equation}
  where $re^{\mp i\theta}\equiv 1 \mp i\omega/\kappa$.
  We notice that the real part of $u_*$ depends on $\hat{\omega}$ and it becomes
  larger as $\hat{\omega}$ increases.
  This implies that the Kruskal modes with very high frequencies
  $\hat{\omega}$ are relevant for late-time features, while relevant
  frequencies $\hat{\omega}$ will change depending on the time of observation.
[See also discussions around Eq.~(\ref{condition2}).]
  Using 
  \begin{equation}
  \begin{aligned}
   \phi(u_*) &=
   \log \kappa 
   + re^{\mp i\theta}\log\left(\frac{ire^{-1\mp i\theta}}{\hat{\omega}}\right),
   \\
   \phi''(u_*) &= 
   -\kappa^2 re^{\mp i\theta},
  \end{aligned}
  \end{equation}
  we have the saddle point approximation of the integral (\ref{eq:integral}) as
  \begin{equation}
    \int^\infty_{-\infty}e^{\phi(u)}\mathrm du
    \simeq e^{\phi(u_*)} 
    \int^\infty_{-\infty}e^{\frac{\phi''(u_*)}{2}(u-u_*)^2}\mathrm du
    = e^{\phi(u_*)}\sqrt{\frac{2\pi}{-\phi''(u_*)}}
    = \kappa
    \left(
     \frac{ire^{-1\mp i\theta}}{\hat{\omega}}
    \right)^{re^{\mp \theta}}
    \sqrt{\frac{2\pi}{re^{\mp i\theta}}},
    \label{eq:saddle_approx}
  \end{equation}
  which becomes a good approximation when $\omega/\kappa\gg 1$.
  This expression derives the Bogoliubov coefficients of 
  Eq.~(\ref{eq:alpha_beta}) given by 
  \begin{equation}
   \frac{|\alpha_{\omega\hat{\omega}}|^2}{|\beta_{\omega\hat{\omega}}|^2}
    \simeq \exp\left(\frac{2\pi\omega}{\kappa}\right).
  \end{equation}
  Thus, we have the familiar result
  of the Hawking temperature to be $T=\kappa/2\pi$
  by the saddle point approximation.

  We could also perform the integral explicitly to obtain
  a well-known exact expression~\cite{Hawking:1974sw} 
  \begin{equation}
   \int^\infty_{-\infty}e^{\phi(u)}\mathrm du
    = \kappa(-i\hat{\omega})^{-re^{\mp i\theta}}
    \Gamma\left(re^{\mp i\theta}\right).
    \label{exact}
  \end{equation}
  Using the Stirling formula of the gamma function 
  \begin{equation}
   \Gamma(1+z) = z \Gamma(z)
    \simeq \sqrt{2\pi z} \left(\frac{z}{e}\right)^z,
  \end{equation}
  we have Eq.~(\ref{eq:saddle_approx}) again from Eq.~(\ref{exact}).
  This result indicates that when the geometric optics approximation is
  satisfied, namely, when $\omega/\kappa \gg 1$, the saddle point approximation
  is valid.
  
  \subsection{Extension to nonstationary spacetime}
\label{dynamical}

  Let us now take a nonstationary spacetime background.
  To probe this spacetime, we use wave packets
  which are localized in both the time and frequency domains rather than
  plane waves spreading over the whole time.
  The wave packet is peaked around a time $u=u_0$ with width $\sim \Delta u$, 
  which are the time and duration of the observation, respectively.
  The Bogoliubov coefficients for a wave packet are obtained by 
  inserting a window function into the integrand of
  Eq.~(\ref{eq:alpha_beta}),
  that is,
  \begin{equation}
   \left.
    \begin{aligned}
     A_{\omega\hat{\omega}}\\
     B_{\omega\hat{\omega}}
    \end{aligned}
   \right\} = \pm 
   \frac{1}{2\pi}
   \sqrt{\frac{\hat{\omega}}{\omega}}
   \int^{\infty}_{-\infty} w_{\Delta u}(u-u_0) \frac{\mathrm dU}{\mathrm du}
   e^{\pm i \omega u - i \hat{\omega} U(u)}\mathrm du,
   \label{AandB}
  \end{equation}
  where $w_{\Delta u}(x)$ is a window function which goes sufficiently
  fast to zero outside the interval $\Delta u$ around $x=0$.
  In addition, we assume that analytic continuation of $w_{\Delta u}(x)$
  is varying slowly at least for $|x|<\Delta u$ in the complex plane.
  Using the Klein--Gordon product, these coefficients are written as 
  $A_{\omega\hat{\omega}} = (\psi_\omega, \bar u_{\hat{\omega}})$ and 
  $B_{\omega\hat{\omega}} = - (\psi^*_\omega, \bar u_{\hat{\omega}})$,
  where we have constructed the wave packet localized around time $u=u_0$ and
  frequency $\omega$ as 
  \begin{equation}
   \psi_\omega(u) = \frac{1}{\sqrt{4\pi\omega}}
    w_{\Delta u}(u-u_0)e^{-i\omega u}.
  \end{equation}
  We note that $A_{\omega\hat{\omega}}$ and $B_{\omega\hat{\omega}}$ satisfy the
  following relation 
  \begin{equation}
   \int^\infty_0\mathrm d\hat{\omega} 
    \left(|A_{\omega\hat{\omega}}|^2 -|B_{\omega\hat{\omega}}|^2 \right)
    = (\psi_\omega, \psi_\omega),
  \end{equation}
  which follows from the completeness of $\{\bar u_{\hat{\omega}}\}$.
  The number density in terms of the wave packet mode $\psi_\omega$ is
  given by 
  \begin{equation}
   \frac{1}{(\psi_\omega, \psi_\omega)}
    \int^\infty_0\mathrm d\hat{\omega} |B_{\omega\hat{\omega}}|^2,
  \end{equation}
  which is roughly the number of particles generated in the frequency band of
  width $\sim 1/\Delta u$ around frequency $\omega$.

  Let us suppose that the evolution of the background spacetime is
  sufficiently slow around $u=u_0$ within time interval $\Delta u$.
  To clarify what slowly evolving is,
  we shall introduce the following quantity 
  \begin{equation}
   \kappa(u) \equiv - \frac{\mathrm d}{\mathrm du} 
    \log \frac{\mathrm dU}{\mathrm du}.
    \label{kappaofu}
  \end{equation}
  If the background
  spacetime is stationary, $\kappa(u)$ becomes a constant and it is
  nothing but the surface gravity of the black hole (Killing) horizon,
  as we have seen in the previous section.  
  Therefore, we may state that the spacetime is evolving slowly when
  $\kappa(u)$ is almost constant during $\Delta u$.
  In that case, we may express $\kappa(u)$ as
  \begin{equation}
   \kappa(u) = \kappa_0 \left[ 1 + \epsilon f(u) \right],
  \end{equation}
  where $f(u)$ is a real function which satisfies 
  $f(u_0)=0$ and $|f(u)|\leq 1$ for  $|u-u_0|<\Delta u$,
  and $\epsilon$ is a small parameter.
  In other words, this assumption implies  
  \begin{equation}
   \frac{\Delta \kappa}{\kappa_0} \leq \epsilon \quad\text{for}\quad
    |u-u_0|<\Delta u,
    \label{Deltakappa}
  \end{equation}
  where $\Delta\kappa \equiv |\kappa(u) - \kappa_0|$.
  For discussions below, we further assume that analytic continuation of $f(u)$ satisfies 
  $|f(u)|\leq 1$ for any complex $u$ such that $|u-u_0|<\Delta u$.

  Then, we may expand $U'(u)$, $U(u)$ and $U''(u)$ with respect to
  $\epsilon$ as
\begin{equation}
\begin{aligned}
U'(u) &= U'_0 \exp\left(-\int_{u_0}^u \kappa(x) \mathrm dx \right)
= U'_0 \exp\left\{
-\kappa_0\left( \delta u + \epsilon g(u) \right)
\right\}
=
U'_0 e^{-\kappa_0 \delta u} 
\left(
1 - \epsilon \kappa_0 g(u)
+ \mathcal{O}(\epsilon^2)
\right)
,
\\
U(u) &= U_0 + \int_{u_0}^u U'(x) \mathrm dx
=
U_0 + U'_0 \left(
\frac{1 - e^{-\kappa_0\delta u}}{\kappa_0}
- \epsilon \kappa_0 \int_{u_0}^u e^{-\kappa_0 (x-u_0)}g(x)\mathrm dx
+ \mathcal{O}(\epsilon^2)
\right),
\\
U''(u) &=
-\kappa(u)U'(u) 
=
-\kappa_0\left( 1 + \epsilon f(u)\right)U'_0 
e^{-\kappa_0 \delta u}\left(
1 - \epsilon \kappa_0 g(u) + \mathcal{O}(\epsilon^2)
\right),
\end{aligned}
\label{Us}
\end{equation}
where we defined $U_0=U(u_0)$, $\delta u = u-u_0$ and
\begin{equation}
 g(u) = \int_{u_0}^{u}f(x)\mathrm dx.
\end{equation}
Note that $|g(u)|\leq |\delta u|$ follows from the assumption 
$|f(u)|\leq 1$ for $|\delta u|<\Delta u$.
We also expand the saddle point $u=u_*$ as  
 \begin{equation}
  u_* = u_*^{(0)} + \epsilon u_*^{(1)} 
  + \mathcal O (\epsilon^2).
 \end{equation}
Then, the equation for the saddle point from Eq.~(\ref{phip}) 
is expanded using Eq.~(\ref{Us}) as
\begin{equation}
0= \phi'(u_*) =
-\kappa_0 re^{\mp i\theta} 
- i\hat{\omega} U'_0 e^{-\kappa_0\delta \uzero}
+ \epsilon\kappa_0 \left[
-f\bigl(\uzero\bigr)
+ i\hat{\omega} U'_0 e^{-\kappa_0\delta \uzero}
\left(
\uone + g\bigl(\uzero\bigr)
\right)
\right]
+ \mathcal{O}(\epsilon^2),
\label{eqsaddle2}
\end{equation}
where we have redefined $re^{\mp i\theta}\equiv 1\mp i\omega/\kappa_0$, 
which satisfies $r \gg 1$ because of the geometric optics approximation
$\omega/\kappa_0\gg 1$.
Solving 
$\phi'(u_*) = 0$ order by order of $\epsilon$, we find 
\begin{equation}
\begin{aligned}
 \delta\uzero 
=&
-\kappa_0^{-1}\log\left(\frac{i\kappa_0re^{\mp i\theta}}{\hat{\omega}U'_0}\right)
= -\frac{1}{\kappa_0}\log\left(\frac{\kappa_0r}{\hat{\omega}U'_0}\right)
 - \frac{i}{\kappa_0}\left(\frac{\pi}{2} \mp \theta\right)
,
\\
\uone =&
- g\bigl(\uzero\bigr) 
+ \frac{e^{\kappa_0\delta\uzero}}{i\hat{\omega}U'_0}f\bigl(\uzero\bigr)
=
- g\bigl(\uzero\bigr) - \frac{f\bigl(\uzero\bigr)}{\kappa_0re^{\mp i\theta}}~.
\end{aligned}
\label{saddle}
\end{equation}
Plugging Eq.~(\ref{saddle}) into Eq.~(\ref{phi})
and $\phi''(u) = -\kappa'(u)-i\hat{\omega} U''(u)$ at $u=u_*$
and expanding with respect to $\epsilon$, we find 
$\phi(u_*) - \phi(u_0) \equiv \phi_0 + \epsilon \phi_1$
and
$\phi''(u_*) \equiv \phi''_0 + \epsilon \phi''_1$
are given by
 $\phi(u_0)\equiv \log U'_0 \pm i\omega u_0-i\hat{\omega}U_0$,
\begin{equation}
 \phi_0 =
  -re^{\mp i\theta}  \left(
		      \kappa_0 \delta\uzero + 1
       \right)
  - \frac{i\hat{\omega}U'_0}{\kappa_0}
  ,
  \qquad
  \phi_1
  = -\kappa_0 g\bigl(\uzero\bigr) 
  + i\hat{\omega}U'_0\kappa_0
  \int_{u_0}^{\uzero} e^{-\kappa_0 \delta u}g(u)\mathrm du
  \label{phiser}
\end{equation}
and
\begin{equation}
\phi''_0 =
-\kappa_0^2 re^{\mp i\theta}
,
\qquad
\phi''_1 =
-\kappa_0\left(
f'\bigl(\uzero\bigr) 
+\kappa_0 \left(1+re^{\mp i\theta}\right) f\bigl(\uzero\bigr)
\right),
\label{phipser}
\end{equation}
where $\mathcal{O}(\epsilon^2)$ terms are omitted.

To guarantee that the perturbative expansion above is valid and that the calculation 
of the saddle point approximation is not affected by the correction terms, we should
require each of correction terms [$\epsilon\uone$, $\epsilon\phi_1$ and 
$\epsilon\phi''_1$ in Eqs.~(\ref{saddle}), (\ref{phiser}) and (\ref{phipser})]
to be much smaller than their leading term ($\delta\uzero$, $\phi_0$ and $\phi''_0$, respectively).
As for $\epsilon\bigl|\uone\bigr|\ll\bigl|\delta\uzero\bigr|$ 
and $\epsilon\bigl|\phi_1''\bigr|\ll\bigl|\phi_0''\bigr|$,
we can see from Eqs.~(\ref{saddle}) and (\ref{phipser}) that we should assume
both $\epsilon \ll 1$ and $\bigl|f'\bigl(\uzero\bigr)\bigr|\ll \kappa_0 r$
to hold to satisfy such requirements, where we used $|g(u)|\leq|\delta u|$.
As for $\epsilon |\mathrm{Re}\,\phi_1|\ll |\mathrm{Re}\,\phi_0|$,
we have 
\begin{equation}
 \begin{aligned}
  \mathrm{Re}\,\phi_0 =& \mp \omega \mathrm{Im}\, u^{(0)}_*
  - \kappa_0 \mathrm{Re}\,\delta u^{(0)}_* - 1,\\
  \mathrm{Re}\,\phi_1 =& - \kappa_0 \mathrm{Re}\,g\bigl(u^{(0)}_*\bigr)
  + \mathrm{Re}\left(
  i \hat{\omega}U'_0\kappa_0 
  \int^{\mathrm{Re}\,u^{(0)}_* + i  \mathrm{Im}\,u^{(0)}_*}_{\mathrm{Re}\,u^{(0)}_*}
  e^{-\kappa_0 \delta u}g(u)\mathrm du
  \right),
 \end{aligned}
\end{equation}
and 
\begin{equation}
 |\mathrm{Re}\,\phi_1| < \kappa_0 \bigl|\delta u^{(0)}_*\bigr|
  + \kappa_0^2 r \bigl|\delta u^{(0)}_*\bigr| \bigl|\mathrm{Im}\,u^{(0)}_*\bigr|.
\end{equation}
Then, for $\epsilon |\mathrm{Re}\,\phi_1|\ll |\mathrm{Re}\,\phi_0|$ to hold,
we need to additionally require $\epsilon\kappa_0\bigl|\delta u^{(0)}_*\bigr|\ll 1$.
Under these assumptions, 
we find that the integral in Eq.~(\ref{AandB}) 
can be evaluated by the saddle point approximation as
  \begin{equation}
   \left|\int^\infty_{-\infty} w_{\Delta u}(\delta u)e^{\phi(u)}\mathrm du\right|
    \simeq U'_0
    \left|w_{\Delta u}\bigl(\delta \uzero\bigr)\right|
    e^{\mathrm{Re}\,\phi_0}\sqrt{\frac{2\pi}{|\phi''_0|}}~.
  \end{equation}
So that the suppression due to the window function, $w_{\Delta u}$, 
is weak and the integral value becomes close to that for the unwindowed function,
 the saddle point should satisfy $|\delta u_*|\lesssim \Delta u$.
  For the imaginary part of $u_*^{(0)}$ the condition 
  $|\delta u_*|\lesssim \Delta u$ becomes
  \begin{equation}
   \frac{1}{\Delta u} \alt \kappa_0,
   \label{condition1}
  \end{equation}
  and for the real part it becomes a condition on $\hat{\omega}$, given by
  \begin{equation}
   \frac{\kappa_0 r}{U'_0}e^{-\kappa_0\Delta u} \alt \hat{\omega} 
    \alt \frac{\kappa_0 r}{U'_0}e^{\kappa_0\Delta u}.
    \label{condition2}
  \end{equation}
  If $\hat{\omega}$ is out of this region, no saddle point 
  exists between $|u-u_0|\alt \Delta u$,
  and then the integral will be suppressed due to the window function.
  This condition means that only the Kruskal modes of $\hat{\omega}$
  with limited frequency band, specified by Eq.~(\ref{condition2}),
  can have correlations with the wave packet mode
  $\psi_\omega$ due to its localization in the time domain.

  Consequently, sufficient conditions for the saddle point
  approximation to be valid are given by
  \begin{equation}
   \epsilon\kappa_0 \ll \frac{1}{\Delta u} \alt \kappa_0 \ll \omega,
    \qquad
    \bigl|f'(u)\bigr|\ll \kappa_0 r,
    \label{condition}
  \end{equation}
  for $|u-u_0| < \Delta u$.
  Note that the condition on $\epsilon$, $\epsilon \kappa_0\ll \Delta u^{-1}$, can 
  be rewritten as $\Delta \kappa \Delta u\ll 1$, where $\Delta \kappa$ is 
  defined by Eq.~(\ref{Deltakappa}).
  If we can take the time interval $\Delta u (\gtrsim \kappa_0^{-1})$
  satisfying this condition, we have a sufficiently small $\epsilon \ll 1$ for
  which Eq.~(\ref{condition}) holds.
  Roughly speaking, 
  this condition is indicating that the time variation of
  $\kappa(u)$ should be sufficiently moderate to satisfy
  $\frac{1}{\kappa_0^2}\frac{\mathrm d\kappa}{\mathrm du} \ll 1$ 
  between the interval 
  $|u-u_0| \lesssim \kappa_0^{-1}$. 
  Under these assumptions, at the leading order we have 
  \begin{equation}
   \frac{|A_{\omega\hat{\omega}}|^2}{|B_{\omega\hat{\omega}}|^2}
    \simeq 
  \exp\left(\frac{2\pi\omega}{\kappa_0}\right),
  \end{equation}
  which implies the spectrum observed at the time $u=u_0$ becomes a thermal one 
  with temperature $\kappa_0/2\pi$ for high frequencies $\omega \gg \kappa_0$.

  \subsection{Surface gravity for past horizon}
  \label{past}  

  In this section, we consider the geometrical meaning of $\kappa(u)$
  defined by Eq.~(\ref{kappaofu}) in the previous section.
    Ingoing null vectors with respect to null
  coordinates $u$ and $U$ are written as 
  \begin{equation}
   k^a = \left(\frac{\partial}{\partial u}\right)^a, \quad
    \bar k^a = \left(\frac{\partial}{\partial U}\right)^a,
  \end{equation}
  respectively.
  Note that $U$ is an affine parameter on the past horizon because it is
  one of the Kruskal coordinates. 
  Using $k^a = U'(u)\bar k^a$, we have 
  \begin{equation}
   k^a \nabla_a k^b = k^a \nabla_a (U'(u)\bar k^b)
    = (k^a \nabla_a U'(u))\bar k^b
    + (U')^2 \bar k^a \nabla_a \bar k^b.
  \end{equation}
  The last term vanishes because $U$ is the affine parameter. 
  As a result, we obtain $k^a\nabla_a k^b = - \kappa(u) k^b$, where 
  \begin{equation}
   \kappa(u) = - \frac{\mathrm d}{\mathrm du} 
    \log \frac{\mathrm dU}{\mathrm du}.
  \end{equation}
  We shall call $\kappa(u)$ ``surface gravity'' for the past horizon
  because it describes the inaffinity of the null generator $k^a$ of the past
  horizon, which is defined by the asymptotic time at the null infinity.
  When the spacetime is stationary, 
  the past horizon and $\kappa(u)$ coincide with the Killing horizon and its
  surface gravity, respectively.
  In this sense,
  $\kappa(u)$ is a natural extension of surface gravity of stationary spacetimes. 
  (In Ref.~\cite{Barcelo:2010xk}, it was called ``peeling
  properties'' for gravitational collapse.)
  It is worth noting that the surface gravity for the past horizon is
  not determined only by local geometrical quantities on that horizon but
  also by the relation between the time coordinates on the horizon and in the asymptotic region.
  In fact, the Killing vector which defines the surface gravity of the Killing horizon 
  is also determined using the asymptotic time in a stationary case.
  
  Roughly speaking, 
  the particles of the Hawking radiation observed in the
  asymptotic region start from the vicinity of the horizon and propagate along
  outgoing null geodesics.
  Every outgoing null ray arriving at the null infinity experiences
  the near-horizon region of the past horizon rather than the (future) event horizon.
  Based on this observation,
  it seems to be natural that the spectrum is affected by the surface
  gravity near the past horizon.

 \section{Applications}
 \label{sec:applications}
 
 In this section, we consider explicit examples of transitions from an
 initial equilibrium state to another equilibrium state.
 First, as a simplest example, we consider null shell accretion into a
 black hole in asymptotically flat spacetime.
 Next, we will focus attention on the AdS--Vaidya spacetime.
 This is a toy model to describe a thermalization process in 
 the AdS/CFT correspondence.

  \subsection{Asymptotically flat case: null shell accretion} 

  In this section, we consider a black hole in asymptotically flat spacetime, 
  whose mass is initially $M_\mathrm{i}$ and changes into the final mass 
  $M_\mathrm{f}$ due to accretion of a null shell.
  Then, the Hawking temperature should initially be that of the initial
  black hole and eventually become the temperature of the final one. 
  The metric of a static and spherically symmetric black hole solution is given by 
  \begin{equation}
   \mathrm ds^2 = - 
   f_\mathrm{I}
   (r)\mathrm dt^2 
    + \frac{\mathrm dr^2}{ 
    f_\mathrm{I}
    (r)}
    + r^2 \mathrm d\Omega^2,
    \label{eq:BH_metric}
  \end{equation}
  where the subscript $I$ is $i$ or $f$, 
  which are for initial and final quantities, respectively.
  Double-null coordinates $(u,v)$ are given by 
  \begin{equation}
   v-u = 2\int\frac{\mathrm dr}{f(r)}, \qquad u+v = 2t
  \end{equation}
  
  To describe null shell accretion, we will join two spacetimes at
  $v=0$ in a manner such that for $v<0$ we use the metric of the initial
  black hole and for $v>0$ that of the final one.
  At $v=0$ the junction condition implies that the radial coordinate
  should be identified as 
  \begin{equation}
   u_\mathrm{i} = -2\int\frac{\mathrm dr}{f_\mathrm{i}(r)}, 
   \qquad
    u_\mathrm{f} = -2\int\frac{\mathrm dr}{f_\mathrm{f}(r)}.
  \end{equation}
  Also, it leads to the following relations 
  \begin{equation}
   \frac{\mathrm dR}{\mathrm du_\mathrm{i}}
   = - \frac{f_\mathrm{i}(R)}{2}, \qquad
   \frac{\mathrm dR}{\mathrm du_\mathrm{f}}
   = - \frac{f_\mathrm{f}(R)}{2},
  \end{equation}
  where $R$ is the radius of the null shell, and these are nothing but
  the equations of motion for the null shell.

  The Kruskal coordinate at the past horizon, namely, that of the initial
  black hole, is 
  \begin{equation}
   U = - \exp(-\kappa_\mathrm{i} u_\mathrm{i}),
  \end{equation}
  where $\kappa_\mathrm{i}$ denotes the surface gravity of the initial
  black hole, 
  and the retarded time at null infinity is that of the final black hole
  as $u=u_\mathrm{f}$.
  
  \begin{equation}
   \frac{\mathrm dU}{\mathrm du}
    = \kappa_\mathrm{i} \exp(-\kappa_\mathrm{i} u_\mathrm{i})
    \frac{\mathrm du_\mathrm{i}}{\mathrm du}
    = \kappa_\mathrm{i} \exp(-\kappa_\mathrm{i} u_\mathrm{i})
    \frac{f_\mathrm{f}(R)}{f_\mathrm{i}(R)}
  \end{equation}
  The surface gravity of the past horizon is 
  \begin{equation}
   \begin{aligned}
    \kappa (u)
    =& - \frac{\mathrm d}{\mathrm du}\log\frac{\mathrm dU}{\mathrm du}
    = \kappa_\mathrm{i}\frac{\mathrm du_\mathrm{i}}{\mathrm du}
    - \frac{f'_\mathrm{f}(R)}{f_\mathrm{f}(R)}
    \frac{\mathrm dR}{\mathrm du}
    + \frac{f'_\mathrm{i}(R)}{f_\mathrm{i}(R)}
    \frac{\mathrm dR}{\mathrm du}\\
    =& \left(\kappa_\mathrm{i} - \frac{f'_\mathrm{i}(R)}{2}\right)
    \frac{f_\mathrm{f}(R)}{f_\mathrm{i}(R)}
    + \frac{f'_\mathrm{f}(R)}{2},
   \end{aligned}
  \end{equation}
  where $R(u)$ denotes the radius of the null shell at time $u$.  
  For early time $u\to-\infty$, the radius of the shell becomes
  $R\to\infty$
  and we have $\kappa(u) \to \kappa_\mathrm{i}$. 
  For late time $u\to\infty$, the radius of the shell approaches the
  horizon radius of the final black hole $R\to r_\mathrm{f}$
  and we have $\kappa(u) \to \kappa_\mathrm{f}$ as $u\to\infty$, where we have
  used $\kappa_\mathrm{f} = f'_\mathrm{f}(r_\mathrm{f})/2$.
  These imply that the asymptotic observers detect a change of the Hawking 
  radiation at the retarded time
  when the null shell comes into the vicinity of the black hole horizon.
  
  Consider the four-dimensional Schwarzschild case with
  $f_\mathrm{I}(r) = 1-2M_\mathrm{I}/r$ for example.
  We have 
  \begin{equation}
   \kappa(u) = \frac{R(u)-2\Delta M}{4M_\mathrm{i}R(u)},\qquad
    \frac{\mathrm d}{\mathrm du}\kappa(u)
    = - \frac{\Delta M (R(u) - 2M_\mathrm{f})}{4M_\mathrm{i}R(u)^3},
  \end{equation}
  where $\Delta M \equiv M_\mathrm{f} - M_\mathrm{i}$.
   If $M_\mathrm{i} \gg \Delta M$, the condition 
  \begin{equation}
   \frac{1}{\kappa^2}\frac{\mathrm d\kappa}{\mathrm du} \alt \frac{\Delta M}{M_\mathrm{i}} \ll 1
  \end{equation}
  is satisfied for any retarded time $u$.
  Therefore, it is concluded that the Hawking temperature observed by
  asymptotic observers at time $u$ is given by $\kappa(u)/2\pi$ in the
  current case. 
  We note that $\kappa(u)$ changes gradually
  even though the spacetime describing the null shell
  accretion is not smooth.

  \subsection{Asymptotically anti-de Sitter case: AdS--Vaidya}

  Next, we discuss an asymptotically anti-de Sitter background.
  In this case, null infinity at which the Hawking temperature would be
  observed is not a null surface but a timelike surface, namely, the so-called
  AdS boundary.  
  Moreover, even when there is no black hole horizon, asymptotically AdS spacetime has 
  a past horizon (not a white hole horizon but a Cauchy horizon).

  We consider the $5$D AdS--Vaidya spacetime
  \begin{equation}
   \mathrm ds^2 = \frac{1}{z^2}
    [-F(\bar v,z)\mathrm d\bar{v}^2 - 2\mathrm d\bar v\mathrm dz
    + \mathrm d{\vec{x}_3}^2],
  \end{equation}
  where $F(\bar v,z)$ is given by 
  \begin{equation}
   F(\bar v,z) = 1 - 2m(\bar v)z^4,
  \end{equation}
  and the curvature radius of the AdS is set to unity.
  We suppose that the mass function $m(\bar v)$ is 
  \begin{equation}
   m(\bar v) = 
    \left\{
     \begin{aligned}
      m_0& \quad &(\bar v<0)\\
      m_0& + \Delta m \sin^2 \frac{\pi \bar v}{2\Delta v} \quad 
      &(0 \le \bar v \le \Delta v)\\
      m_0& + \Delta m \quad &(\bar v>\Delta v)
     \end{aligned}
    \right. ~.
  \end{equation}
  During an interval $\Delta v$, the null fluid is injected into the
  black hole with mass $m_0$, and the mass eventually becomes 
  $m_0 + \Delta m$. 
  For the AdS--Vaidya, Bondi energy observed at the boundary is
  described by the mass function $m(\bar v)$.
  Note that the time coordinate $\bar v$ is nothing but asymptotic time
  at the boundary.
  In the context of the AdS/CFT correspondence, $m(\bar v)$ corresponds to the energy
  density of the CFT matter.
  Therefore, the duration of the injection $\Delta v$ will represent
  the time scale of energy density change. 

  Now we shall consider the Hawking temperature in the current case.
  We introduce double-null coordinates $(u,v)$ as 
  \begin{equation}
   z = z(u,v), \quad \bar v = v.
  \end{equation}
  In these coordinates, the boundary $z=0$ is described by $u=v$. 
  An asymptotic time $t$ at the boundary is given by $t=\frac{u+v}{2}$,
  and it is equal to $v$ (or $u$) there.
  The coordinate condition for $(u,v)$
  to be the double-null coordinates leads to  
  \begin{equation}
   \frac{\partial z}{\partial v} = - \frac{F(\bar v,z)}{2},
    \label{eq:dzdv}
  \end{equation}
  which is equivalent to the geodesic equation for the outgoing null ray
  described by $u$-constant line.
  Note that we have 
  $\left.\frac{\partial z}{\partial v}\right|_{z=0} = -
  \left.\frac{\partial z}{\partial u}\right|_{z=0} = -\frac{1}{2}$
  as the boundary condition at $z=0$.
  
  In order to calculate the surface gravity for the past horizon we need
  to know the relation between the asymptotic time and the affine
  parameter at the past horizon. 
  Because the spacetime is static (strictly speaking, independent of $\bar v$) for the
  initial black hole region ($\bar v<0$), the canonical null coordinate
  $\bar u$ is given by 
  \begin{equation}
   \mathrm d\bar u = d\bar v + \frac{2}{F}\mathrm dz,
  \end{equation}
  where we note that the metric function $F$ depends only on $z$.
  Moreover, the affine parameter at the past horizon, namely, the Kruskal
  coordinate $U$ is immediately given by 
  $U(\bar u) = - \exp(-\kappa_\mathrm{i} \bar u)$,
  where $\kappa_\mathrm{i}$ is the surface gravity determined by the
  initial black hole with the mass $m_0$. 
  It is worth noting that, in general, the above null coordinate $\bar u$ is
  different from the asymptotic time $u$ defined previously.
  Let us recall the definition of $u$.
  The differential form becomes 
  \begin{equation}
   \mathrm dz = \frac{\partial z}{\partial u}\mathrm du
    + \frac{\partial z}{\partial v}\mathrm dv
    = - \frac{F}{2}(-\lambda \mathrm du + \mathrm dv),
  \end{equation}
  where $\lambda$ is defined by 
  \begin{equation}
   \lambda(u) \equiv \frac{2}{F}\frac{\partial z}{\partial u}
    = \frac{\mathrm d\bar u}{\mathrm du}.\label{eq:red_shift}
  \end{equation}
  If the spacetime is static, we can take the canonical double-null form
  such that $\lambda = 1$ everywhere.
  It turns out that $\lambda$ describes the relation between the
  asymptotic time $u$ at the boundary and the canonical time $\bar u$
  in the initial black hole region, that is, the redshift factor for
  the outgoing null ray which goes from the initial black hole region to
  the boundary, due to dynamical background.
  (See the Appendix for details.)
  
  Now, we can obtain $\lambda(u)$ as follows.
  Differentiating Eq.~(\ref{eq:dzdv}) with respect to $u$, we have 
  \begin{equation}
   \frac{\partial}{\partial v} \frac{\partial z}{\partial u}
    = - \frac{1}{2}\frac{\partial F}{\partial z} 
    \frac{\partial z}{\partial u}.
  \end{equation}
  Integrating the above equation and the geodesic equation with respect to
  $v$ from the boundary $v=u$ to the initial static region ($v<0$), 
  we obtain $\frac{\partial z}{\partial u}$
  for $v<0$.
  Note that we must integrate those to the past horizon ($v=-\infty$) in
  general.
  However, in the current case it is enough to evaluate only up to
  $v=0$, 
  because the spacetime becomes static for $v<0$ and the redshift
  factor $\lambda(u)$ does not change any more.

  As a result, we have 
  \begin{equation}
   \kappa(u) = - \frac{\mathrm d}{\mathrm du}
    \log \frac{\mathrm dU}{\mathrm du}
    = \kappa_\mathrm{i} \lambda(u)
    - \frac{\mathrm d}{\mathrm du}\log\lambda(u),
  \end{equation}
  where we have used 
  $\frac{\mathrm dU}{\mathrm du} = \lambda \frac{\mathrm dU}{\mathrm d\bar u}$.
  
\begin{figure}
   \begin{center}
    \begin{tabular}{cc}
     \resizebox{80mm}{!}{\includegraphics{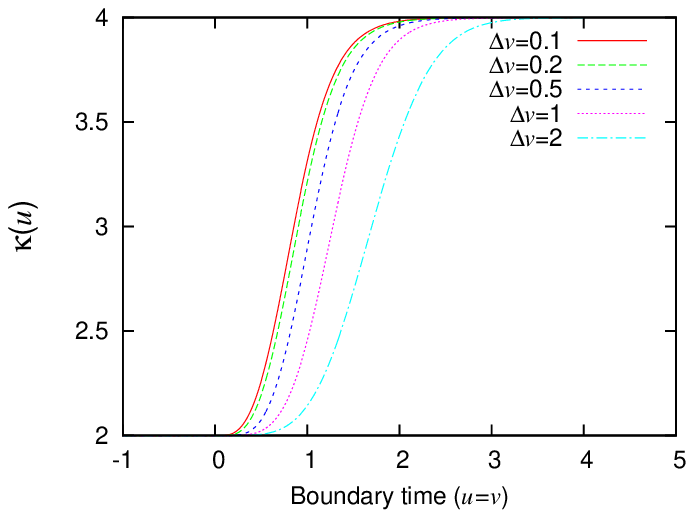}} &
     \resizebox{80mm}{!}{\includegraphics{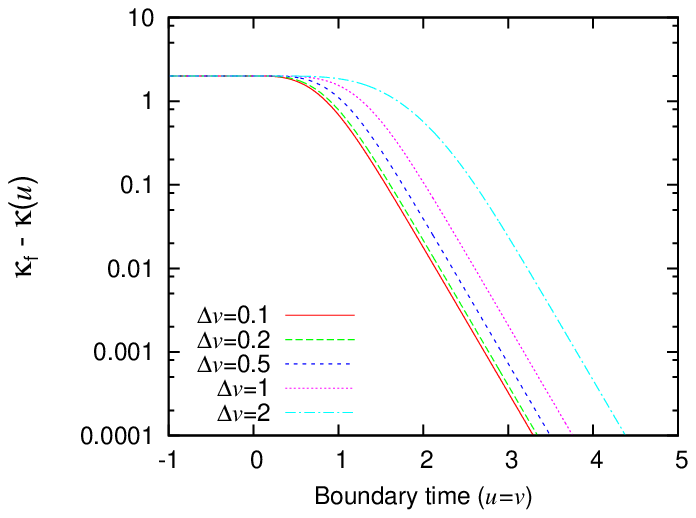}}
    \end{tabular}
    \caption{
    $\kappa(u)$ as a function of time ($u=v$) at the boundary (left) and
    behaviors of $\kappa(u)$ near the final temperature
    $\kappa_\mathrm{f}$ (right). 
    Time scales of injection are taken to be 
    $\Delta v = 0.1, 0.2, 0.5, 1, 2$, which are shorter than a
    typical time scale given by temperature.
    We set $m_0 = 1/2$ and $\Delta m = 15/2$,
    which give $\kappa_\mathrm{i}=2$ and $\kappa_\mathrm{f}=4$, respectively.}
    \label{fig:short}
   \end{center}
  \end{figure}

\begin{figure}
   \begin{center}
    \resizebox{80mm}{!}{\includegraphics{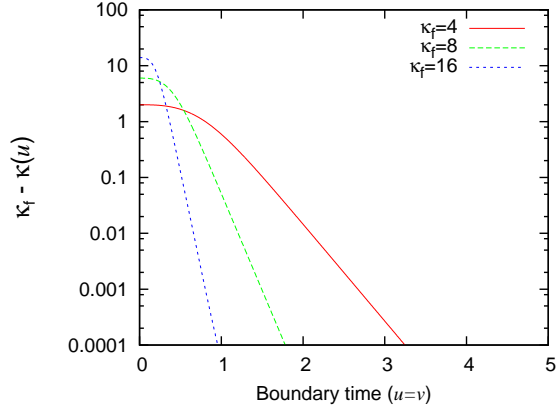}}
    \caption{
    Relaxation rate for final temperatures $\kappa_\mathrm{f} = 4,8,16$ 
    for the case of null shell accretion ($\Delta v=0$).
    The initial temperature is given by $\kappa_\mathrm{i} = 2$.
    In any case,
    an exponential decay phase starts after a transient phase,
    and the decay rate in the exponential decay phase is proportional to 
    inverse of the final temperature.
    }
    \label{fig:rate}
   \end{center}
  \end{figure}

  In Fig.~\ref{fig:short} we plot $\kappa(u)$ as a function of the
  boundary time ($u=v$) for shorter injection times
  $\Delta v = 0.1, 0.2, 0.5, 1, 2$.
  The initial and final temperatures are given by $\kappa_\mathrm{i} = 2$ 
  and $\kappa_\mathrm{f} = 4$, respectively.
  The conditions of Eq.~(\ref{condition}) are almost satisfied
  for these parameters
  and thus the results of the saddle point approximation are
  valid.\footnote{
  If the difference between the initial and final temperatures
  becomes very large, the adiabatic condition may be
  temporarily violated. This means that the system is so far from
  equilibrium that temperature can not be defined at that time. However,
  after the system has relaxed near the final temperature, 
  the interpretation of $\kappa(u)$ as temperature becomes well-defined again.}
  In these cases, we find that 
  $\kappa(u)$ converges into the final temperature exponentially after 
  a transient phase.
  Time scales of the whole process of temperature change are
  roughly given by $\sim 3$ and do not strongly depend on time scales of
  injection but are dominated by exponential relaxation to the final temperature. 
  Recalling that the time scale determined by temperature $T$ of a black hole is
  given by $1/T=2\pi/\kappa$, we may interpret this result as the time scale of
  the temperature change is governed by the temperature of the final
  black hole.
  See also Fig.~\ref{fig:rate} about final-temperature dependence of
  relaxation rate for the case of null shell accretion,
  for which the injection is instantaneous and equivalent to the case of 
  $\Delta v =0$.
Note that since we are focusing on time scales shorter than the thermal time scale 
$\Delta u \sim \kappa^{-1}$ and the minimum time resolution of observation is
given by this thermal time scale,
we should interpret $\kappa(u)$ not as a value at
time $u$ but rather a time-averaged value over time scale $\sim \kappa^{-1}$.
  
\begin{figure}
   \begin{center}
    \begin{tabular}{cc}
     \subfigure[$\Delta v=0$]{\resizebox{60mm}{!}{\includegraphics{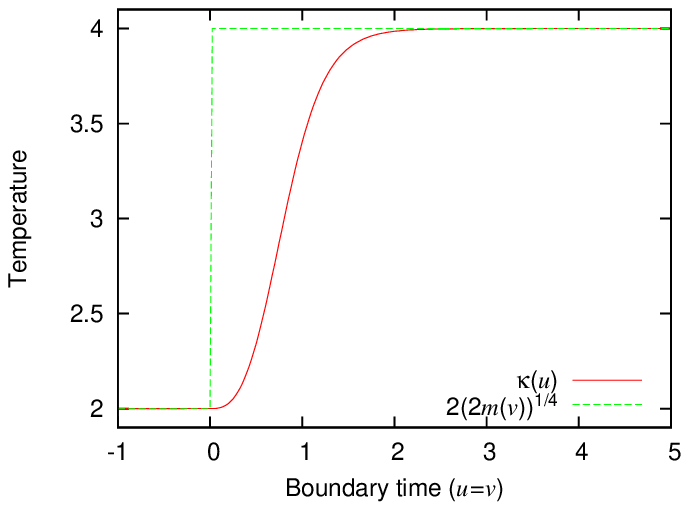}}} &
     \subfigure[$\Delta v=2$]{\resizebox{60mm}{!}{\includegraphics{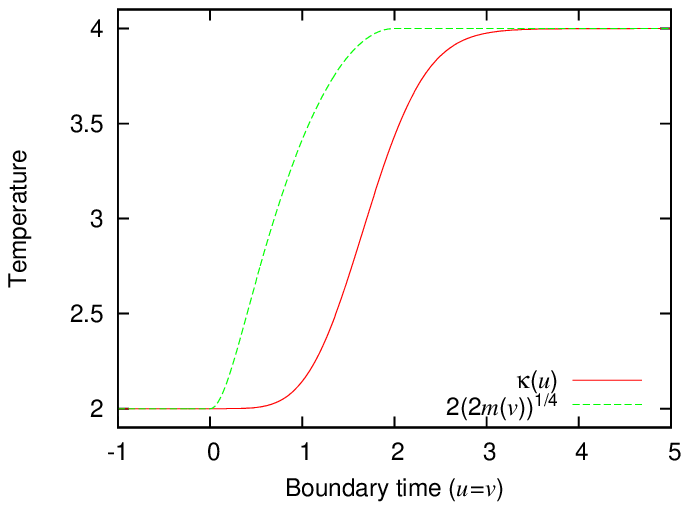}}} \\
     \subfigure[$\Delta v=10$]{\resizebox{60mm}{!}{\includegraphics{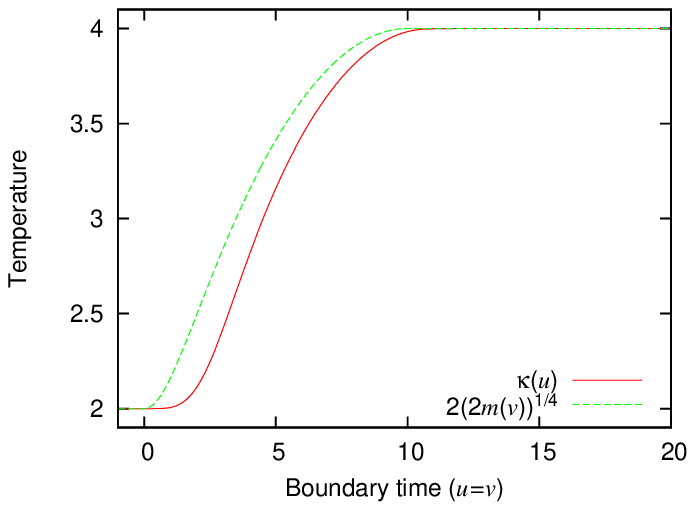}}} &
	 \subfigure[$\Delta v=50$]{\resizebox{60mm}{!}{\includegraphics{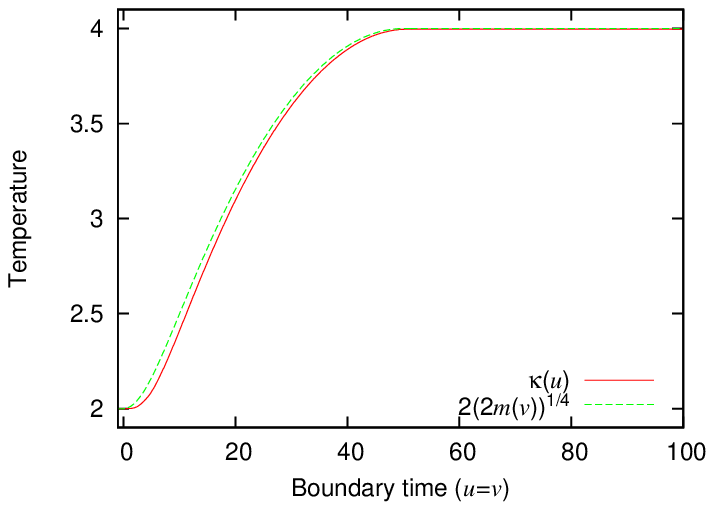}}}
    \end{tabular}
    \caption{$\kappa(u)$ and a quasistatic temperature 
    $2\pi T_\mathrm{qs} = 2(2m(v))^{1/4}$ for various injection times 
    $\Delta v$.
    }
    \label{fig:long}
   \end{center}
  \end{figure}

As the injection time becomes longer than the thermal time scale,
$\kappa(u)$ begins to show qualitatively new behaviors.
  In Fig.~\ref{fig:long}, for $\Delta v = 0, 2, 10, 50$
  we plot $\kappa(u)$ together with 
  a quasistatic temperature $2\pi T_\mathrm{qs}=2\left(2m(v)\right)^{1/4}$, 
  which is naively determined from the mass functions $m(v)$ in a
  similar manner to static black holes.
  We can observe that both curves begin to coincide as the injection time becomes longer.
  In other words, the variation of the temperature $\kappa(u)$ tends to
  follow the variation of the black hole mass $m(v)$.
  
  These results lead us to the conclusion that, even if the mass
  injection is rapid, the temperature will gradually change over a time scale
  no shorter than the thermal time scale $\sim\kappa^{-1}$.
  In this sense,
  we may say that there is a finite relaxation time irrespective of 
  the speed of energy injection.
  When energy injection and the resultant black hole evolution 
  is slower than this relaxation time,
  the temperature tends to evolve in a similar way to the mass injection.

 \section{Summary and Discussion}
 \label{Sec:summary}
 
 In this paper we have discussed Hawking temperature for 
 nonstationary spacetimes.
 We introduced a useful measure for time evolution of spacetime, 
 and showed that the Hawking temperature at each time is determined by 
 the surface gravity of the past horizon when the time evolution 
 of the spacetime is sufficiently slow.
 This definition of temperature is quite natural because 
 observers can determine it without using 
 causally inaccessible information,
 such as the position of the event horizon.

 Although we have considered an eternal black hole which has the past
 horizon, the essential calculation does not change for
 black hole formation in an asymptotically flat spacetime. 
 In this case
 we should draw null rays from future null infinity back to 
 the past null infinity instead of the past horizon,
 and as a result
 we can reproduce the original derivation of the Hawking radiation.
 The definition of $\kappa(u)$ is the same as Eq.~(\ref{kappaofu}),
 even though it cannot be interpreted as ``surface gravity'' and 
 should be understood rather in the context of ``peeling properties'' 
 of outgoing null geodesics, which was discussed in~\cite{Barcelo:2010xk}. 
 It would be interesting to study the properties of $\kappa(u)$ for
 black holes in an expanding universe, such as those discussed in 
 Refs.~\cite{Saida:2007ru,Kaloper:2010ec}, for which 
 the derivation is similar to the case of asymptotically flat spacetime.
 In the case of asymptotically AdS spacetime, 
 in  the Poincar\'e chart 
 the past horizon, which
 is the Cauchy horizon, exists even in the pure AdS spacetime.
 This setup is quite similar to the eternal black hole, and we can naturally 
 interpret $\kappa(u)$ as the ``surface gravity'' of this past horizon.

 In the context of the AdS/CFT correspondence,
 our results indicate that quantum fluctuation of the boundary CFT is governed
 by the surface gravity of the past horizon when the bulk is the eternal
 black hole, namely, the thermal equilibrium state. 
 Particularly, our result can be applied to
 phenomena discussed in Refs.~\cite{deBoer:2008gu,Aharony:2005bm},
 which are directly related to the Hawking radiation in the bulk.
 It may be interesting to study its implications further.
 It would be also interesting to study $\kappa(u)$ in numerical solutions
 obtained in, e.g., Refs.~\cite{Chesler:2008hg,Murata:2010dx,Garfinkle:2011hm}, though
 in general numerical calculation of $\kappa(u)$ becomes difficult at late time
 due to exponential pileup of outgoing null geodesics onto the horizon.

 It is shown that the change of the Hawking temperature is delayed
 by thermal time scale
 compared with the mass injection from the boundary.
 This behavior can be interpreted as follows:
 The injected matter falls down into the black hole and changes the
 geometry in the vicinity of the horizon.
 After that, the Hawking radiation affected by this change emanates 
 from there and reaches the boundary.
 This process of infall of the matter and return of the modified radiation
 will take finite time and, roughly speaking, this is the origin of the
 delay we observed.
 From the viewpoint of the boundary theory, 
 this delay time can be interpreted as the relaxation time scale 
 needed to achieve thermal equilibrium after the energy injection.

 For planar AdS black holes (namely, large black holes in asymptotically
 AdS spacetime) the horizon is located relatively near to the AdS boundary.
 If a black hole evolves sufficiently slower than the delay time, we
 may neglect its delay.
 In that case, the quasistatic approximation, such that one synchronizes
 dynamics of the event or apparent horizon and the boundary in terms of the advanced
 time and uses quantities associated with the future horizon, 
 might be justified.

 What we emphasize is that the ``surface gravity'' of the past horizon,
 $\kappa(u)$, is not only a geometrical quantity but also one with a physical
 meaning in the sense that it governs the thermal spectrum of the Hawking
 radiation observed by asymptotic observers.
 Many of thermodynamic properties of black holes, however, tend to be
 associated with the future event or apparent horizon in the previous works, not
 with the past horizon.
 It is open to discussion how these points of view are related to each other.
 It may be also fruitful to study how $\kappa(u)$ is related to other probes of the
 black hole spacetime, such as those discussed in
 Ref.~\cite{Balasubramanian:2011ur}, and how our approach is related to other
 derivations of the Hawking radiation, such as the tunneling approach.

 \begin{acknowledgments}
  We would like to thank Takahiro~Tanaka for helpful discussions.
  We would also like to thank Nemanja~Kaloper, McCullen~Sandora, 
  Ken-ichi~Nakao, Shinji~Mukohyama and Tadashi~Takayanagi for valuable comments.
  N.T.~acknowledges hospitality at the Centro de Ciencias de Benasque during the
  Strings and Gravity Workshop, and thanks the participants for useful discussions.
  S.K.~is supported by JSPS Grant-in-Aid for Creative Scientific Research No.~19GS0219.
  N.T.~is supported in part by the DOE Grant DE-FG03-91ER40674.
 \end{acknowledgments}

 \appendix
 
 \section{redshift factor}
 \label{app:red_shift}
 
 In this appendix we show that $\lambda(u)$ defined by
 Eq.~(\ref{eq:red_shift}) is the redshift factor for the outgoing null ray.
 Now, we recall that the two-dimensional part of the metric is given by 
 \begin{equation}
  \mathrm ds^2 = \frac{1}{z^2}[-F(\bar v,z)\mathrm d\bar v^2
   - 2 \mathrm d\bar v \mathrm dz],
 \end{equation}
 A coordinate transformation such as 
 $z=z(u,v)$ and $\bar v=v$
 to the double-null coordinates $(u,v)$ gives us the differential form  
 \begin{equation}
  \mathrm d z = \frac{\partial z}{\partial u}\mathrm d u
   + \frac{\partial z}{\partial v}\mathrm d v
   = - \frac{F}{2}(-\lambda \mathrm d u + \mathrm d v),
 \end{equation}  
 where we have defined as 
 $\lambda \equiv - \left.\frac{\partial z}{\partial u}\right/
 \frac{\partial z}{\partial v}$.  
 
 We consider a timelike vector field $\xi^a = \partial/\partial \bar{v}$ which
 characterizes a natural Killing time in stationary regions and also in
  asymptotic regions near the boundary.
  Note that frequencies of the Killing modes are defined with respect to
  this Killing time.
  By using the $1$-form it is rewritten as 
  \begin{equation}
   \xi_a = - \frac{1}{z^2}(F\mathrm d\bar{v} + \mathrm dz).
  \end{equation}
  The tangent vector of outgoing null geodesics is given by 
  \begin{equation}
   l^a \equiv \frac{\mathrm d}{\mathrm ds},
  \end{equation}
  where $s$ is an affine parameter, and then the geodesic equations lead to 
  \begin{equation}
   \frac{\mathrm d}{\mathrm ds}\frac{\dot{\bar v}}{z^2}
    = \frac{\partial F}{\partial z} \frac{{\dot{\bar v}}^2}{2z^2},\qquad
    \dot z = - \frac{F}{2}\dot{\bar v},
  \end{equation}
  where the dot denotes the derivative with respect to $s$.
  We note that the last equation is equivalent to the coordinate
  condition $\partial z/\partial v = - F/2$ defining the
  double-null coordinates.

  The ``Killing energy'' $E$ associated with the null geodesics is defined by 
  \begin{equation}
   E \equiv - l^a \xi_a
    = \frac{1}{z^2}(F\dot{\bar v} + \dot z)
    = \frac{F}{2}\frac{\dot{\bar v}}{z^2}.
  \end{equation}
  If $\xi^a$ is truly a Killing field, $E$ should be constant along the
  null geodesics.
  From the geodesic equations we have 
  \begin{equation}
   \frac{\mathrm d}{\mathrm d\bar v} \frac{\dot{\bar v}}{z^2}
    = \frac{1}{2}\frac{\partial F}{\partial z}
    \frac{\dot{\bar v}}{z^2},
  \end{equation}
  where we have used 
  $\mathrm d/\mathrm ds = \dot{\bar v} \,\mathrm d/\mathrm d\bar v$.
  On the other hand, differentiating  
  $\partial z/\partial v = - F/2$ with respect to $u$, we have 
  \begin{equation}
   \frac{\partial}{\partial v} \frac{\partial z}{\partial u}
    = - \frac{1}{2}\frac{\partial F}{\partial z}
    \frac{\partial z}{\partial u}.
  \end{equation}
  As a result, we find 
  \begin{equation}
   \frac{\partial z}{\partial u} = C\frac{z^2}{\dot{\bar v}},
  \end{equation}
  where $C$ is an integration constant which can be absorbed into
  normalization of the affine parameter.
  Then, the ``Killing energy'' can be rewritten as 
  \begin{equation}
   E = - C\frac{\partial z}{\partial v}
    \left/\frac{\partial z}{\partial u}\right. .
  \end{equation}
  Since the boundary conditions at $z=0$ are given by 
  \begin{equation}
   \left.\frac{\partial z}{\partial v}\right|_{z=0}
    = -\left.\frac{\partial z}{\partial u}\right|_{z=0}
    = -\frac{1}{2},
  \end{equation}
  we have $E_\mathrm{b} = C$, which is the energy observed at the
  boundary, and 
  \begin{equation}
   E|_{z=z(s)} = E_\mathrm{b}/\lambda.
  \end{equation}

  Using the geodesic equations, we also have
  \begin{equation}
   \frac{\mathrm dE}{\mathrm ds} = 
    \frac{F}{2}\frac{\partial F}{\partial z} \frac{{\dot{\bar v}}^2}{2z^2}
    + \frac{\dot{\bar v}}{2z^2}\frac{\mathrm dF}{\mathrm ds}
    = \frac{\dot{\bar v}}{2z^2}\dot{\bar v}
    \frac{\partial F}{\partial \bar v}
    = E \dot{\bar v} \frac{\partial \ln F}{\partial \bar v}.
  \end{equation}
  Integrating it along the outgoing null geodesics described by
  $z=z_\mathrm{g}(\bar v)$, we obtain  
  \begin{equation}
   E_\mathrm{b} = E_\mathrm{i}
    \exp \int_{\bar{v}_\mathrm{i}}^{\bar{v}_\mathrm{b}} \mathrm d\bar v 
    \left(\frac{\partial \ln F}{\partial \bar v}
    \right)_{z=z_\mathrm{g}(\bar v)},
  \end{equation}
  where $E_\mathrm{i}$ and $E_\mathrm{b}$ are the Killing energy
  observed at an initial surface and the boundary, respectively.
  
  Consequently, for the outgoing null ray described by
  $u=\mathrm{const.}$, we have the redshift factor from an initial time
  $\bar{v}_\mathrm{i}$ to asymptotic time $u = \bar{v}_\mathrm{b}$ as 
  \begin{equation}
   \lambda(u) =  
    \exp \int_{\bar{v}_\mathrm{i}}^{u} \mathrm d\bar v 
    \left(\frac{\partial \ln F}{\partial \bar v}
    \right)_{z=z_\mathrm{g}(\bar v)}.
  \end{equation}
  It turns out that if the spacetime is stationary, namely, $F$ does
  not depend on $\bar v$, we have $\lambda=1$.
  The redshift factor $\lambda(u)$ 
  deviates from the unity when time dependence is turned on.


\begin{thebibliography}{99}

\bibitem{Hawking:1974sw}
  S.~W.~Hawking,
  Commun.\ Math.\ Phys.\  {\bf 43}, 199-220 (1975).

\bibitem{Maldacena:1997re}
  J.~M.~Maldacena,
  Adv.\ Theor.\ Math.\ Phys.\  {\bf 2}, 231 (1998)
  [Int.\ J.\ Theor.\ Phys.\  {\bf 38}, 1113 (1999)]
  [arXiv:hep-th/9711200].

\bibitem{Witten:1998zw}
  E.~Witten,
  Adv.\ Theor.\ Math.\ Phys.\  {\bf 2}, 505 (1998)
  [arXiv:hep-th/9803131].


\bibitem{Harada:2000ar}
  T.~Harada, H.~Iguchi and K.~i.~Nakao,
  Phys.\ Rev.\  D {\bf 62}, 084037 (2000)
  [arXiv:gr-qc/0005114].

\bibitem{Saida:2007ru}
  H.~Saida, T.~Harada and H.~Maeda,
  Class.\ Quant.\ Grav.\  {\bf 24}, 4711 (2007)
  [arXiv:0705.4012 [gr-qc]].

\bibitem{Nielsen:2007ac}
  A.~B.~Nielsen, J.~H.~Yoon,
  Class.\ Quant.\ Grav.\  {\bf 25}, 085010 (2008).
  [arXiv:0711.1445 [gr-qc]].

\bibitem{Hayward:2008jq}
  S.~A.~Hayward, R.~Di Criscienzo, L.~Vanzo, M.~Nadalini and S.~Zerbini,
  Class.\ Quant.\ Grav.\  {\bf 26}, 062001 (2009)
  [arXiv:0806.0014 [gr-qc]].

\bibitem{Barcelo:2010pj}
  C.~Barcelo, S.~Liberati, S.~Sonego and M.~Visser,
  Phys.\ Rev.\  D {\bf 83}, 041501 (2011)
  [arXiv:1011.5593 [gr-qc]].

\bibitem{Barcelo:2010xk}
  C.~Barcelo, S.~Liberati, S.~Sonego and M.~Visser,
  JHEP {\bf 1102}, 003 (2011)
  [arXiv:1011.5911 [gr-qc]].

\bibitem{Mazumder:2011gk}
  N.~Mazumder, R.~Biswas and S.~Chakraborty,
  arXiv:1106.4375 [gr-qc].





\bibitem{Son:2002sd}
  D.~T.~Son, A.~O.~Starinets,
  JHEP {\bf 0209}, 042 (2002).
  [hep-th/0205051].



\bibitem{Janik:2006gp}
  R.~A.~Janik, R.~B.~Peschanski,
  Phys.\ Rev.\  {\bf D74}, 046007 (2006).
  [hep-th/0606149].


\bibitem{Kinoshita:2008dq}
  S.~Kinoshita, S.~Mukohyama, S.~Nakamura, K.~-y.~Oda,
  Prog.\ Theor.\ Phys.\  {\bf 121}, 121-164 (2009).
  [arXiv:0807.3797 [hep-th]].

\bibitem{Chesler:2008hg}
  P.~M.~Chesler and L.~G.~Yaffe,
  Phys.\ Rev.\ Lett.\  {\bf 102}, 211601 (2009)
  [arXiv:0812.2053 [hep-th]].

\bibitem{Murata:2010dx}
  K.~Murata, S.~Kinoshita, N.~Tanahashi,
  JHEP {\bf 1007}, 050 (2010).
  [arXiv:1005.0633 [hep-th]].

\bibitem{Das:2010yw}
  S.~R.~Das, T.~Nishioka and T.~Takayanagi,
  JHEP {\bf 1007}, 071 (2010)
  [arXiv:1005.3348 [hep-th]].

\bibitem{Hubeny:2010ry}
  V.~E.~Hubeny and M.~Rangamani,
  Adv.\ High Energy Phys.\  {\bf 2010}, 297916 (2010)
  [arXiv:1006.3675 [hep-th]].

\bibitem{Ebrahim:2010ra}
  H.~Ebrahim and M.~Headrick,
  arXiv:1010.5443 [hep-th].

\bibitem{Hashimoto:2010wv}
  K.~Hashimoto, N.~Iizuka and T.~Oka,
  Phys.\ Rev.\  D {\bf 84}, 066005 (2011)
  [arXiv:1012.4463 [hep-th]].

\bibitem{Erdmenger:2011jb}
  J.~Erdmenger, S.~Lin and T.~H.~Ngo,
  JHEP {\bf 1104}, 035 (2011)
  [arXiv:1101.5505 [hep-th]].

\bibitem{CaronHuot:2011dr}
  S.~Caron-Huot, P.~M.~Chesler, D.~Teaney,
  Phys.\ Rev.\  {\bf D84}, 026012 (2011).
  [arXiv:1102.1073 [hep-th]].

\bibitem{Balasubramanian:2011ur}
  V.~Balasubramanian {\it et al.},
  Phys.\ Rev.\  D {\bf 84}, 026010 (2011)
  [arXiv:1103.2683 [hep-th]].

\bibitem{Garfinkle:2011hm}
  D.~Garfinkle, L.~A.~Pando Zayas,
  Phys.\ Rev.\  {\bf D84}, 066006 (2011).
  [arXiv:1106.2339 [hep-th]];
 D.~Garfinkle, L.~A.~P.~Zayas, D.~Reichmann,
  [arXiv:1110.5823 [hep-th]].




\bibitem{deBoer:2008gu}
  J.~de Boer, V.~E.~Hubeny, M.~Rangamani and M.~Shigemori,
  JHEP {\bf 0907}, 094 (2009)
  [arXiv:0812.5112 [hep-th]].

\bibitem{Aharony:2005bm}
  O.~Aharony, S.~Minwalla, T.~Wiseman,
  Class.\ Quant.\ Grav.\  {\bf 23}, 2171-2210 (2006).
  [hep-th/0507219].

\bibitem{Klemm:1998bb}
  D.~Klemm and L.~Vanzo,
  Phys.\ Rev.\  D {\bf 58}, 104025 (1998)
  [arXiv:gr-qc/9803061].

\bibitem{Hemming:2000as}
  S.~Hemming, E.~Keski-Vakkuri,
  Phys.\ Rev.\  {\bf D64}, 044006 (2001).
  [gr-qc/0005115].


\bibitem{Kaloper:2010ec}
  N.~Kaloper, M.~Kleban, D.~Martin,
  Phys.\ Rev.\  {\bf D81}, 104044 (2010).
  [arXiv:1003.4777 [hep-th]].


 \end{thebibliography}
\end{document}